\documentclass[%
reprint,
nofootinbib,
amsmath,
amssymb,
preprintnumber,
]{revtex4}

\usepackage{graphicx}% Include figure files
\usepackage{dcolumn}% Align table columns on decimal point
\usepackage{bm}% bold math
\usepackage[dvipsnames]{xcolor}

\begin{document}

\preprint{CYCU-HEP-21-02}

\title{Dressed Tunneling in Soft Hair}%

\author{Wen-Yu Wen}\thanks{wenw@cycu.edu.tw}
 \affiliation{Department of Physics and Center for Theoretical Physics,\\
 Chung Yuan Christian University, Taoyuan, Taiwan}
 \affiliation{Leung Center for Cosmology and Particle Astrophysics, National Taiwan University, Taipei, Taiwan}

\begin{abstract}
We revisit the Parikh-Wilczek's tunneling model of Hawking radiation in the Schwarzschild black hole with soft hair. Unlike the no-hair black hole, tunneling through the degenerate event horizon generated by supertranslation hair picks up an angular dependence, which suggests a {\sl hairy} modification to the Bekenstein-Hawking entropy.  At last, we discuss several implications and possible observable of soft hair. 
\end{abstract}


\maketitle

%\tableofcontents

\section{Introduction}
Parikh and Wilczek earlier gave a semiclassical derivation of Hawking radiation as a tunneling process, similar to pair creation in a constant electric field \cite{Parikh:1999mf,Srinivasan:1998ty}.  If one considers a particle with energy $\omega$ emitted from a black hole of mass $M$, the emission rate reads
\begin{equation}\label{eqn:tunnel_rate}
\Gamma \sim e^{-2 \text{Im} S} = e^{\Delta S_{BH}}
\end{equation}
Here the natural unit is adopted that $\hbar=c=1$ and Boltzmann constant $k_B=1$.  If causing no confusion, we also set $G=1$ or equivalently the Planck length $l_p=1$ for convenience.  It is impressive that a nonthermal spectrum can be obtained by the assumption that energy is conservation through the tunneling process.  In addition, the exponent happens to be the very change of Bekenstein-Hawking entropy  $\Delta S_{BH}=4\pi(M-\omega)^2-4\pi M^2$, conservation of information is therefore achieved in terms of mutual information \cite{Zhang:2009jn}.  Nevertheless, it is unclear whether the infamous paradox of lost information can be resolved in this macroscopic picture since it still cannot reveal those hidden microstates responsible for the black hole entropy.   Bondi, van der Burg, Metzner and Sachs (BMS) independently demonstrated the spacetime has an infinitesimal dimensional group associated with asymptotic symmetries \cite{Bondi:1962,Sachs:1962zza,Sachs:1962}.  Weinberg later found S-matrix element of n-particles could relate to one another with additional zero 4-momentum soft photon or graviton, which plays a role in removal of infrared divergence in quantum field theory \cite{Weinberg:1965nx}.   Later, Strominger and his collaborators verified that the soft graviton theorem is exactly equivalent to the Ward identity of BMS supertranslation and superrotation, and their relation to the memory effect \cite{Strominger:2013jfa, He:2014laa, Strominger:2014pwa}.  This impressive triangular relation among BMS symmetry, soft theorem and memory effect has shed new insight into black hole physics.  Hawking, Perry and Strominger noticed an infinite family of degenerate vacua associated with BMS supertranslation at null infinity.  They suggested an infinite numbers of soft hairs of black hole are responsible for storage of were-claimed-lost information \cite{Hawking:2016msc,Strominger:2017aeh}.  Recently, it was shown that soft hair model could have resolved the AMPS (Almheiri-Marolf-Polchinski-Sully) firewall paradox \cite{Almheiri:2012rt,Pasterski:2020xvn} and reproduced the Page curve for unitary evolution \cite{Cheng:2020vzw}.

\section{Tunneling through degenerate horizon}

The tunneling rate (\ref{eqn:tunnel_rate}) is independent on the choice of coordinates, for examples, in the Gullstrand-Painlev\'{e} coordinates \cite{Parikh:1999mf} or the isotropic coordinates \cite{Wang:2007zzm}.  In partuicular, the spatial part of isotropic metric is conformally flat and there is equal chance for particles to tunnel outwards as antiparticles to tunnel inwards.  Isotropic metric has the advantage to describe a static spherically symmetric perfect fluid, widely used in modeling compact stellar objects such as white dwarfs and neutron stars.  As a star collapses to form a black hole, information (or entropy) is somehow encoded in the soft hair state $|C\rangle$ labeled by some function $C$.  It is convenient to adopt same metric form at this transition.  On the other hand, as a black hole radiates, stored information is possibly carried away via the featured Hawking radiation, along with transition between different soft hair states, say $|C\rangle \to |C^\prime\rangle$.  The isotropic metric would be a superior choice to those soft hair states because it faithfully measures the angle at constant time hyperslices and function $C$ can be decomposed into its spherical harmonic components.

Under supertranslation, the hairy Schwarzschild black hole in isotropic coordinate becomes\cite{Compere:2016hzt}:
\begin{widetext}
\begin{equation}\label{eqn:hair_metric}
ds^2=-\frac{(1-\frac{M}{2\rho_s})^2}{(1+\frac{M}{2\rho_s})^2}dt^2+(1+\frac{M}{2\rho_s})^4\bigg( d\rho^2+\big(\big((\rho-E)^2+U \big)\gamma_{AB} + (\rho-E)C_{AB}\big) dz^A dz^B \bigg),
\end{equation}
\end{widetext}
where scalars $E, U$ and tensor $C_{AB}$ are functions of $C(z^A)$, and their precise forms are irrelevant here except to note that $\rho_s = \sqrt{(\rho-C-C_{0,0})^2+||{\cal D} C||^2}$.  Here $C_{0,0}$ refers to the zero mode in spherical harmonic expansion, which generates a time translation, and the square of norm $||{\cal D} C||^2 \equiv \gamma_{AB}D^AC D^BC$.  The metric reduces to the no-hair Schwarzschild solution for vanishing $C$.  Otherwise, the horizon will be deformed by $C$ and some explicit examples were given in \cite{Lin:2020gva,Takeuchi:2021ibg}.  Rather than S-wave tunneling in \cite{Parikh:1999mf}, we assume tunneling a mass $\omega$ through the horizon with specific soft hair along a fixed direction.  The imaginary part of action for outgoing particle at semi-classical approximation becomes\cite{Wang:2007zzm}:
\begin{equation}
    \text{Im} S_{out} = Im \int_0^{\omega}\int_{\rho_{in}}^{\rho_{out}}\frac{(\rho_s+\frac{(M-\omega^\prime)}{2})^3}{(\rho_s-\frac{(M-\omega^\prime)}{2})\rho_s^2}d\rho d(-\omega^\prime).
\end{equation}
With a change of variable $d\rho = \frac{\rho_s}{\sqrt{\rho_s^2-||{\cal D}C||^2}} d\rho_s$, one can evaluate the integral by deforming the contour around the single pole at $\rho_s=\frac{M-\omega^\prime}{2}$ and obtain:
\begin{widetext}
\begin{equation}\label{eqn:action_out}
    \text{Im} S_{out} = \pi\bigg \{(M-\omega^\prime)\sqrt{(M-\omega^\prime)^2-4||{\cal D}C||^2}+4||{\cal D}C||^2\ln\big\{ (M-\omega^\prime)+\sqrt{(M-\omega^\prime)^2-4||{\cal D}C||^2} \big\} \bigg\}_0^\omega
\end{equation}
\end{widetext}
Similarly for incoming antiparticle, one evaluates $\text{Im} S_{in}=-\text{Im} S_{out}$ and the net $\text{Im} S = \text{Im} S_{out}- \text{Im} S_{in} = 2 \text{Im} S_{out}$.  This gives us the tunneling rate per solid angle:
\begin{equation}
 \frac{d\Gamma}{d\Omega} = \frac{1}{2\pi}\text{Im} S_{out}.   
\end{equation}
In the case of no hair $C=0$, $-\text{Im} S_{out} = 2\pi M^2 - 2\pi (M-\omega)^2 = \Delta S/2$ as expected in the \cite{Parikh:1999mf}.

\section{Quantum-corrected entropy}
According to the Parikh-Wilczek tunneling model, the imaginary action (\ref{eqn:action_out}) suggests a {\sl hairy} form for the Bekenstein-Hawking entropy {\sl density}, that is\footnote{A nonzero constant term can be added to (\ref{eqn:nonlinear_correction}), which serves as the UV cut in relation to the remnant \cite{Chen:2009ut}.  In the section VII, we will argue this constant term is not needed in our construction.}

\begin{widetext}
\begin{equation}\label{eqn:nonlinear_correction}
    \frac{dS}{d\Omega} = M\sqrt{M^2-4||{\cal D}C||^2} + 4 ||{\cal D}C||^2 \ln \big\{ M+\sqrt{M^2-4||{\cal D}C||^2} \big\}
\end{equation}
\end{widetext}
It is expected that (\ref{eqn:nonlinear_correction}) does not respect the area law thanks to the deformed horizon.  To make sense of this nonlinear form, one can expand it for large $M$ or small $||{\cal D}C||^2$:
\begin{equation}\label{eqn:linear_correction}
    S = 4\pi M^2 + \underbrace{16\pi \overline{||{\cal D}C||^2} \ln M}_{\text{logarithmic correction}}  -\frac{41}{2}\pi\overline{||{\cal D}C||^4} \frac{1}{M^2} + \cdots,
\end{equation}
where $\overline{||{\cal D}C||^2}$ is the angular average of $||{\cal D}C||^2$. The logarithmic and other corrections to Bekenstein-Hawking entropy have been widely discussed; for example, see \cite{Page:2004xp} for a review. as well as its connection to tunneling method  \cite{Vanzo:2011wq}.  We have following remarks. Firstly, the correction terms in (\ref{eqn:linear_correction}) are similar to those in \cite{Banerjee:2008ry}, which involve a logarithmic term and all powers of inverse area ${\cal O}(M^{-2n})$.  In some sense, one may regard the degenerate horizon as a kind of {\sl locally} modified surface gravity.  Secondly, it was found earlier that the coefficient in front of each correction term is either a constant \cite{Kaul:2000kf,Carlip:2000nv} or spin-dependent \cite{Fursaev:1994te}.  In our construction, the coefficient uniquely depends on nonvanishing norm $||{\cal D}C||$.  In fact, it only permits a positive coefficient thanks to space-like vector ${\cal D}C$.  Thirdly, those terms with square root and logarithm in  (\ref{eqn:nonlinear_correction}) will remain real and positive as long as $M$ is large, which is justified by the semi-classical approximation.  Validation of (\ref{eqn:nonlinear_correction}) beyond semi-classical limit will be discussed in the section VII.      At last, although the zero mode $C_{0,0}$ does not have effect on the logarithmic correction, it may still incur a linear correction to the Bekenstein-Hawking entropy while tunneling through a Vaidya black hole \cite{Chu:2018tzu}.

\section{Surface gravity at deformed horizon}
The hair metric (\ref{eqn:hair_metric}) is obtained from the Schwarzschild metric via supertranslation diffeomorphism\cite{Compere:2016hzt}.  The surface gravity at the Schwarzschild coordinate $\rho_s=M/2$ is a constant as expected for a static background.  However, since the Parikh-Wilczek tunneling is a dynamic process and out of thermal equilibrium, an uneven gravity or temperature gradient at the horizon might be closer to the reality.   We take the near-horizon limit of hair metric (\ref{eqn:hair_metric}), say $\rho_s \simeq \frac{M}{2}+Mx$ for $x \ll 1$:
\begin{equation}
 ds^2 \simeq \underbrace{-x^2dt^2 + \kappa^{-2} dx^2}_{\text{Rindler space}} + \underbrace{\cdots}_{\text{Sphere part}},  
\end{equation}
where we identify the surface gravity $\kappa=\frac{\sqrt{M^2-4||{\cal D}C||^2}}{4M^2}$ and define the hair temperature \`{a} la Unruh:
\begin{equation}{\label{eqn:hair_temp}}
    T_{hair}=\frac{\kappa}{2\pi}.
\end{equation}
Later, we will regard this hair temperature as the non-equilibrium temperature $\Theta$.  Our viewpoint is as follows: each solid angular patch reaches quasi-thermal equilibrium at hair temperature.  Since the hair temperature is angle dependent, its gradient is perpendicular to the radial direction.  The uneven surface gravity might induce tidal force on pair produced particles.  Their transverse motion can be seen as heat flow driven by the temperature gradient. We will elaborate this point in the section VIII.

\section{Cosmic censorship}
The conjecture of cosmic censorship implies that $\frac{M^2}{4}-||{\cal D}C||^2 \ge 0$, otherwise the singularity would be naked \cite{Compere:2016hzt}.  In fact, this inequality can be also obtained from the reality condition of  (\ref{eqn:nonlinear_correction}). In the following, we will discuss possible implication of this conjecture.  Firstly, we assume that censorship is still protected after the black hole emits a particle of mass $\omega>0$ and hair function changes from $C\to C'$.  Provided that $C \sim {\cal D}C \sim \epsilon M$, the inequality of censorship directly translates into the inequality between the change of entropy $\Delta S_{BH}$ and that of hair function $\Delta(||{\cal D}C||^2)$:
\begin{equation}
\Delta(||{\cal D}C||^2) \le \frac{\Delta S_{BH}}{16\pi} < 0,
\end{equation}
which in turn gives lower bound of tunneling rate: 
\begin{equation}
   e^{16\pi \Delta(||{\cal D}C||^2)} \le \Gamma < 1 
\end{equation}
It is obvious that the inequality of cosmic censorship stands for small perturbation when the black hole is massive.   However, it is nontrivial to validate this conjecture near the extreme limit when $\frac{M}{2}-||{\cal D}C||\equiv \delta$ is a small positive value.  This happens at late stage of evaporation when the black hole is small and light.  We follow the Gedanken experiment first illustrated by Wald \cite{Wald:1974,Wald:1997wa} to test the cosmic censorship by throwing in a test mass $\omega$ into a soft-hair black hole of mass $M$.  If the black hole were destroyed, meaning that the inequality were violated afterwards, then we have the joint inequality:
\begin{equation}
    ||{\cal D}C||  + \delta + \frac{\omega}{2} = \frac{M+\omega}{2} < ||{\cal D}C'||.
\end{equation}
Together we have
\begin{equation}{\label{eqn:WCCC}}
    \omega < 2(\Delta(||{\cal D}C||) - \delta).
\end{equation}
On the other hand, the total mechanical energy for a non-relatvistic particle to reach horizon and form a bound state with the black hole is given by  $E_{\text{mech}}=-\frac{M\omega}{2r_s}\big|_{r_s=2M} \simeq -\frac{\omega}{4}$. Therefore, a negative-mass particle with mass bounded by (\ref{eqn:WCCC}) could never get close to the horizon if it were released from rest at the infinity, no mention to destroy the black hole.  This argument is always valid for a no-hair black hole ($C=0, \delta = M/2$).   However, in the case of soft-hair, additional condition $\Delta(||{\cal D}C||) \le \delta$ is needed to ensure a negative mass in (\ref{eqn:WCCC}).  That is, to ensure the weak consmic censorship conjecture, the change of hair function is bounded by the non-extremal parameter $\delta$.  We remark that Wald's argument is usually trivial for the  Schwarzschild-like black hole with single conserved charge.  Here the soft hair complicates the situation and the censorship conjecture constraints the amount of change in the hair function.  In addition, the above argument is limited to a non-relativistic absorption, a thorough analysis on relativistic energy condition is waiting for future justification.

\begin{figure}
\includegraphics[scale=1]{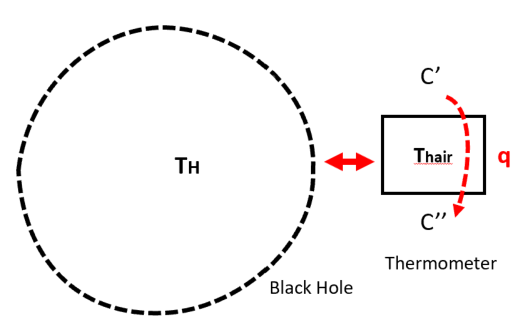}
\caption{\label{fig:thermometer}A Gedanken experiment of thermometer placed near a hairy black hole to measure non-equilibrium temperature. Although the heat exchange between the black hole and the thermometer is balanced, $T_{hair}$ is different from the equilibrium Hawking temperature $T_H$ due to the transverse heat flow $q$ driven by the difference of hair functions $C'$ and $C''$ across the thermometer.}
\end{figure}

\begin{figure}[t]
\includegraphics[scale=0.30]{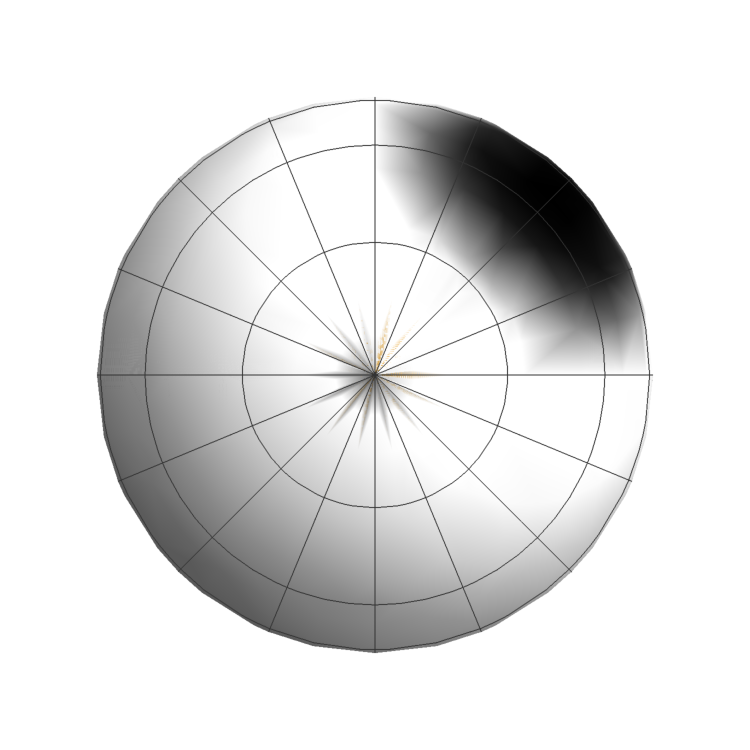}
\includegraphics[scale=0.30]{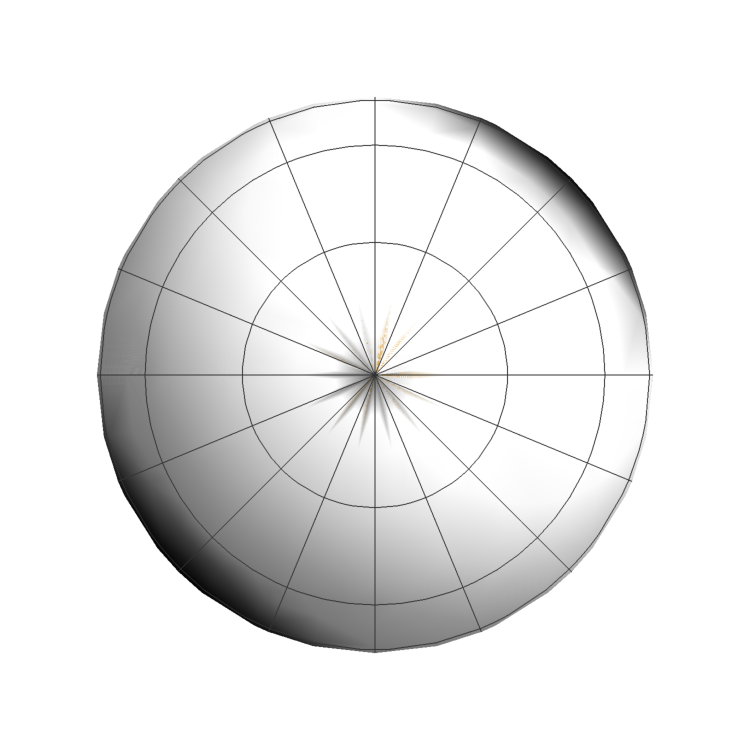}
\includegraphics[scale=0.30]{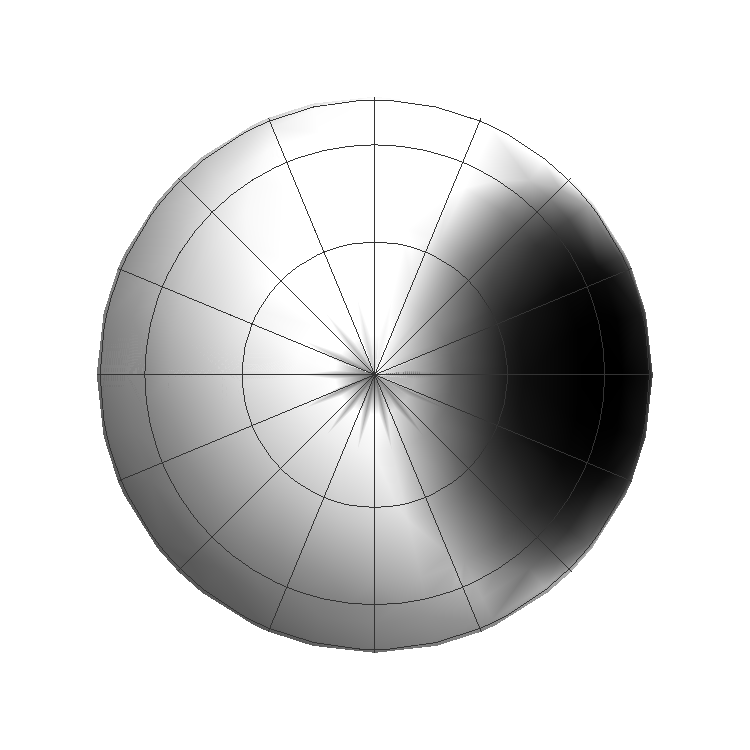}
\includegraphics[scale=0.30]{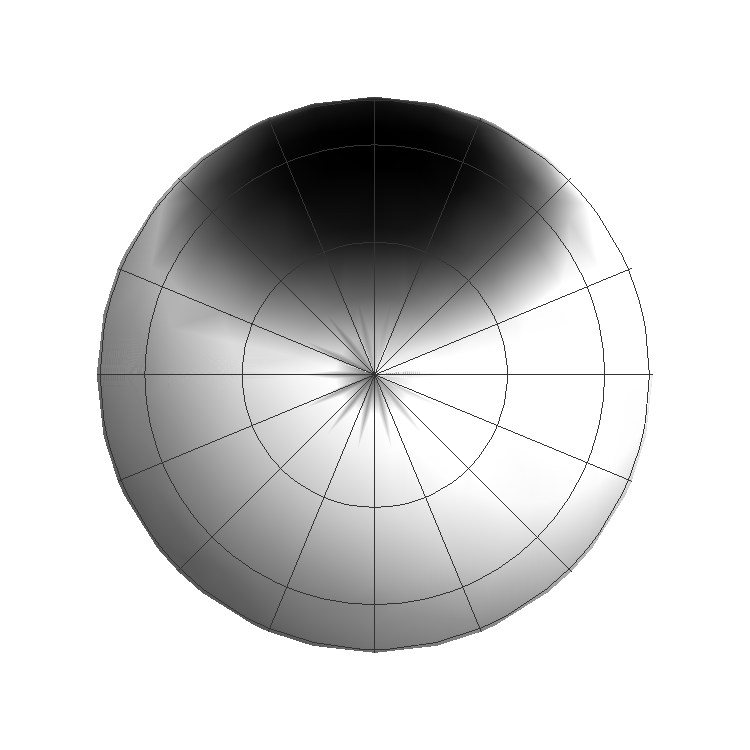}
\caption{\label{fig:observable}Topview of hair temperature fluctuation at event horizon for soft hair function $C\simeq \epsilon M \{ c_{-1}Y_1^{-1}+c_0Y_1^0+c_1Y_1^1 \}$. Coefficients  $(c_{-1},c_0,c_1)$ are chosen as $(1+i,1,-1+i)$ (top left), $(1,1,-1)$ (top right), $(1+i,0,-1+i)$ (bottom left), $(i,1,i)$ (bottom right).  Darker region indicates higher temperature.  Here we only show its relative value with respect to the Hawking temperature, since we do not know the magnitude of  $\epsilon$.}
\end{figure}

\section{Information loss paradox}
If one denotes $\Gamma(M;\omega)$ as the tunneling rate for a black hole of initial mass $M$ emitting a particle of energy $\omega$.  It has been argued in \cite{Medved:2005vw,Braunstein:2011gz} that  the productive relation $\ln[\Gamma(M;\omega_1)\Gamma(M-\omega_1;\omega_2)]=\ln[\Gamma(M;\omega_1+\omega_2)]$ is generally true thanks to conservation of energy and entropy (information).  This implies that the inclusion of quantum correction, such as (\ref{eqn:nonlinear_correction}) or (\ref{eqn:linear_correction}), is not sufficient to late-time correlation, no mention to resolve the information loss paradox.  In other words, the Parikh-Wilczek tunneling model may have claimed that information is not lost, nevertheless it cannot explain how is information encoded or decoded.  However, we emphasize that the hairy correction  (\ref{eqn:nonlinear_correction}) does not satisfy this productive relation due to its nontrivial angular dependency, that is
\begin{equation}\label{eqn:violate_product}
\ln[\Gamma(M;\omega_1;\Delta C_1)\Gamma(M-\omega_1;\omega_2;\Delta C_2)] \neq \ln[\Gamma(M;\omega_3;\Delta C_3)],
\end{equation}
where $\omega_1+\omega_2=\omega_3$.  We highlight the change of soft hair $\Delta C_i$ associated with each emission $\omega_i$.  Note that $\Delta C_3$ needs not to be related to $\Delta C_1$ or $\Delta C_2$ because mass and hair (information) are not simply related, in addition one can always find a different combination such that $\omega_1^\prime+\omega_2^\prime=\omega_3$ to which another $\Delta C_i^\prime$ are associated.  Violation (\ref{eqn:violate_product})  suggests that it is possible to entangle those soft hair states which are responsible for late-time correlation.

\section{Remnants}
If both conservation of entropy and logarithmic correction persists in whole life of black holes, there would be remnants at the end of evaporation process \cite{Chen:2009ut}.  It is still under debate whether remnants can provide a satisfying answer to information loss paradox; see \cite{Chen:2014jwq} for a recent review.  Nevertheless, since $||{\cal D}C||^2 \propto M^2$ by simple dimension argument, the hairy entropy (\ref{eqn:nonlinear_correction}) smoothly goes to zero as $M\to 0$, therefore no place for a remnant in this construction.

\section{Non-equilibrium observables}
Noticing that the Parikh-Wilczek tunneling process is nonthermal in general, where thermal quantities such as entropy or temperature, can only be defined locally in each hair patch.  It is nontrivial and nonuniversal to extend those quantities to describe the whole black hole system.  For our purpose, we find the approach of non-equilibrium thermodynamics adopted in \cite{Casas:2003} become handy and insightful.   We first expand the non-equilibrium entropy $S^{neq}$ around its equilibrium value $S^{eq}$ and identify with those terms in (\ref{eqn:linear_correction}):
\begin{equation}
    S^{neq} = \underbrace{4\pi M^2}_{S^{eq}} + \underbrace{\alpha(M) ||{\cal D}C||^2}_{\text{non-equilibrium perturbation}} + \cdots,
\end{equation}
where the equilibrium entropy $S^{eq}$ is given by the Bekenstein-Hawking entropy.   This formula suggests the equilibrium state is driven away by the non-equilibrium variable (here is the hair function $||{\cal D}C||$) in the quadratic form, where the coefficient $\alpha(M)=16\pi \ln M$ is regarded as the Hessian $\partial^2 S^{neq}/\partial ||{\cal D}C||^2$.  Then the non-equilibrium temperature $\Theta$ can be obtained {\`a} la the first law as follows:
\begin{equation}
    \frac{1}{\Theta} = \frac{\partial S^{neq}}{\partial M} = \frac{1}{T_H} + \frac{16\pi}{M} ||{\cal D}C||^2 + \cdots
\end{equation}
We remark that the non-equilibrium temperature $\Theta$ agrees with $T_{hair}$ upon Taylor expansion at small $||{\cal D}C||/M$.  We also recall that the hair function $C \sim ||{\cal D}C|| \sim {\cal O}(\epsilon)$, therefore we may rewrite 
\begin{eqnarray}\label{eqn:temperature_fluctuation}
    S^{neq} = 4\pi M^2 + {\cal O}(\epsilon^2), \nonumber\\
    \Theta = \frac{1}{8\pi M}+ {\cal O}(\epsilon^2),
\end{eqnarray}
which implies that the non-equilibrium entropy and temperature agree with those of Schwarzschild black holes up to order ${\cal O}(\epsilon^2)$.  
This correction cannot be observed by a detector at infinity if it were in thermal equilibrium with radiation from the Schwarzschild black hole.  Nevertheless, a Gedanken experiment, similar to the design in \cite{Casas:2003}, might illustrate the possibility to measure the fluctuation of hair temperature.  A thermometer of proper size is placed at an infinitesimal distance $\epsilon^2M$ outside the horizon as shown in the FIG. \ref{fig:thermometer}.  Even though the heat exchange (via radiation) between the black hole and the thermometer is balanced, heat flow $q^A$ driven by the hair function gradient $D^AC$ can still flow in transverse direction, which gives rise to the non-equilibrium temperature $T_{hair}$.    In the FIG. \ref{fig:observable}, we sketch the hair temperature fluctuation at the event horizon for various combination of first spherical harmonics.

\begin{acknowledgments}
The author is grateful to Dr. Sayid Mondal and Dr. Shingo Takeuchi for useful discussion.  This work is supported in part by the Taiwan's Ministry of Science and Technology (grant No. 106-2112-M-033-007-MY3 and 109-2112-M-033-005-MY3) and the National Center for Theoretical Sciences (NCTS).

\end{acknowledgments}

\end{document}